\documentclass[aps]{revtex4}
\usepackage{eurosym}
\usepackage{amsfonts}
\usepackage{amsmath}
\usepackage{amssymb,epsf}
\usepackage{color}
\usepackage{graphicx}
\usepackage{epstopdf}
\usepackage{float}
\usepackage{caption}
\usepackage{subfig}

\begin{document}

\title{Three-dimensional accelerating AdS black holes in $F(R)$ gravity}
\author{B. Eslam Panah$^{1,2,3}$\footnote{%
eslampanah@umz.ac.ir}, M. Khorasani$^{4}$\footnote{%
mkhorasani@shirazu.ac.ir}, and J. Sedaghat$^{4}$\footnote{%
J.sedaghat@shirazu.ac.ir}}
\affiliation{$^{1}$ Department of Theoretical Physics, Faculty of Science, University of
Mazandaran, P. O. Box 47416-95447, Babolsar, Iran\\
$^{2}$ ICRANet-Mazandaran, University of Mazandaran, P. O. Box 47416-95447,
Babolsar, Iran\\
$^{3}$ ICRANet, Piazza della Repubblica 10, I-65122 Pescara, Italy\\
$^{4}$ Department of Physics, School of Science, Shiraz University, Shiraz
71946-84795, Iran}

\begin{abstract}
Considering a three-dimensional $C-$metric, we obtain the exact accelerating
black holes in the $F(R)$ theory of gravity coupled with and without a
matter field. First, we extract uncharged accelerating AdS black hole
solutions in $F(R)$ gravity. Then, we study the effects of various
parameters on metric function, roots, and the temperature of these black
holes. The temperature is always positive for the radii less than $\frac{1}{%
\alpha }$, and it is negative for the radii more than $\frac{1}{\alpha }$.
We extend our study by coupling nonlinear electrodynamics as a matter filed
to $F(R)$ gravity to obtain charged black holes in this theory. Next, we
evaluate the effects of different parameters such as the electrical charge,
accelerating parameter, angular, $F(R)$ gravity, and scalar curvature on the
obtained solutions, roots, and temperature of three-dimensional charged
accelerating AdS black holes. The results indicate that there is a root in
which it depends on various parameters. The temperature of these black holes
is positive after this root.
\end{abstract}

\maketitle

\section{Introduction}

The accelerating black hole has some exciting properties among other black
holes. The spacetime of the accelerating black holes is extracted from $C-$%
metric \cite{Kinnersley1970,Plebanski1976,Griffiths2006}. A conical deficit
angle along one polar axis of this black hole provides the force driving
acceleration. To understand them, one can imagine that something similar to
the $C-$metric can describe a black hole accelerated by an interaction with
a local cosmological medium. The asymptotic behavior of the accelerating
black holes described by $C-$metric depends on various parameters, which
lead to an accelerating horizon and complicate the asymptotic structure. In
this regard, some interesting properties of the (un)charged accelerating
(non)rotating (A)dS black holes have been studied in Refs. \cite%
{AC1,AC3,AC4,AC5,AC6,AC7,AC8,AC9,AC11,AC12,AC13,AC14,AC15,AC16}.

The classical electromagnetic theory of Maxwell is a widely used fundamental
theory. However, a singularity at the position of the point charge leads to
an infinite self-energy, which is a fundamental problem of Maxwell's theory.
To overcome this problem, Born and Infeld introduced a nonlinear
electromagnetic (NED) field in 1934 \cite{BornII}. Then, numerous models of
NED were developed (see Refs. \cite{NED1,NED2,NED3,NED4,NED5}, for more
details). Power-Maxwell invariant (PMI) theory is one of these special
classes of NED \cite{NED2}, in which its Lagrangian is an arbitrary power of
Maxwell Lagrangian \cite{PMI}. PMI reduces to a linear Maxwell field with
considering a special value, and also it can remove the singularity at the
position of the point charge \cite{PMII,PMIII}. There is an interesting
property for PMI, and it is related to conformal invariance. In other words,
the energy-momentum tensor of PMI is traceless when the power of the Maxwell
invariant is a quarter of spacetime dimensions, which leads to conformal
invariance. In this case, PMI is known as conformal invariant Maxwell (CIM).
This idea is related to taking advantage of the conformal symmetry to
construct the analogs of the four-dimensional Reissner-Nordstr\"{o}m black
holes with an inverse square electric field in arbitrary dimensions \cite%
{NED2,PMI}.

On the other hand, the string theory partition function contains
higher-curvature terms in the IR limit such that considering Kaluza--Klein
reductions can lead to NED coupling to the Einstein sector \cite{Gibbons2011}%
. In addition, nonlinear kinetic terms for the vector fields also appear in
the quantization of string actions \cite{Fradkin1985}. So, NED theories
introduced as an alternative to Maxwell electrodynamics to solve problems in
quantum electrodynamics, can be considered in the duality and allows
extending the possible holographic field theories. Moreover, there are
interesting motivations of NED for applications in gravity, cosmology, and
CMT as well \cite{Sorokin2022}. For example; i) NED coupled to gravity have
been used to construct regular black hole solutions \cite%
{regular1,regular2,regular3,regular4,regular5,regular6}, ii) to create
cosmological models, which might avoid initial Big-Bang singularities \cite%
{BigB1,BigB2}, iii) to mimic dark energy and generate an accelerated
expansion of the universe \cite{NonUni1,NonUni2,NonUni3}, iv) AdS/CFT
correspondence and gravity/CMT holography \cite%
{AdSCMT1,AdSCMT2,AdSCMT3,AdSCMT4,AdSCMT5}. Also, the effects of NED on the
thermodynamics properties of black holes have been studied in some
literature \cite{NEDBH1,NEDBH2,NEDBH3,NEDBH4,NEDBH5,NEDBH6,NEDBH7}.

An accelerated universe expansion was confirmed by various observational
evidence, for example, the luminosity distance of Supernovae type Ia.
Identifying the cause of this late-time acceleration is a challenging
problem from the cosmology point of view. Among candidates that can explain
it, $F(R)$ gravity as a modified theory of gravity attracted significant
attention because of specific properties. $F(R)$ theory of gravity is the
straightforward modification of Einstein's gravity, which can describe some
phenomena from various points of view. The gravitational action in this
theory is a general function of $R$ \cite{F(R)I,F(R)II,F(R)III,F(R)IV}. This
theory may give a unified description of early-time inflation \cite%
{InflationI,InflationII}. It may describe the structure formation of the
Universe without considering dark energy or dark matter. Moreover, $F(R)$
gravity coincides with Newtonian and post-Newtonian approximations \cite%
{Capozziello2005,Capozziello2007}. It is worth mentioning that this theory
of gravity may extract the whole sequence of the Universe's evolution
epochs: inflation, radiation/matter dominance, and dark energy.

The conjectured equivalence of string theory on AdS spaces and
super-conformal gauge theories living on the boundary of AdS has led to an
increasing interest in asymptotically AdS black holes \cite{AdS1,AdS2,AdS3}.
This interest is based on the fact that the classical super-gravity black
hole solution can furnish important information on the dual gauge theory in
the large N limit (in which N denotes the rank of the gauge group). A
standard example of this is the well-known Schwarzschild AdS black hole,
whose thermodynamics was studied by Hawking and Page in Ref. \cite%
{HawkingP1983}. They discovered a phase transition from thermal AdS space to
a black hole phase, as the temperature increases. The Hawking-Page phase
transition was then reconsidered by Witten in the spirit of the AdS/CFT
correspondence \cite{Witten1998}. In this regard, the study of the
thermodynamics of AdS black holes has been extended in various theories of
gravity with linear and nonlinear Maxwell fields \cite%
{ThAdS1,ThAdS2,ThAdS3,ThAdS4,ThAdS5,ThAdS6,ThAdS7,ThAdS8,ThAdS9,ThAdS10,ThAdS11}%
.

Until now, all extant $C-$metrics describing accelerating black holes are
obtained in the context of Einstein's gravity for three and four dimensions
of spacetime with or without linear and nonlinear electrodynamics fields.
For example, the asymptotically AdS accelerated black holes and their
thermodynamic properties in the presence of the ModMax electrodynamics have
been studied in Ref. \cite{Barrientos2022}. This research opened a window
towards studying radiative spacetimes in NED regimes, as well as raised new
challenges for their corresponding holographic interpretation \cite%
{Barrientos2022}. Also, by considering this NED (MadMax) a novel $C-$metric
which describes accelerated AdS black holes is given in Ref. \cite{Hale2023}%
. On the other hand, the investigations on Einstein's gravity solutions in
three-dimensional spacetime provide a powerful background to study the
physical properties of gravity and examine its viability in lower spacetime
dimensions. Especially, it can be simulated as a toy model of quantum
gravity. This property was discovered after the arguments presented
connecting the possible links between three-dimensional gravitation and the
Chern-Simons theory \cite{Achucarro1986,Witten1986}. Considering Einstein's
gravity, the uncharged accelerating black holes in three-dimensional
spacetime have been extracted in Refs. \cite%
{BTZAcI,BTZAcII,BTZAcIIb,BTZAcIII}. The first solution of the accelerating
BTZ black hole is given by Astorino in Ref. \cite{BTZAcI}. Then, Xu et al.
studied it with more detail in Refs. \cite{BTZAcII,BTZAcIIb}. In this
regard, Arenas-Henriquez et al. evaluated the accelerating systems in $(2+1)-
$dimensional spacetime, and found three exciting classes of geometry by
studying holographically their physical parameters \cite{BTZAcIII}.
Moreover, by coupling Einstein's gravity and CIM, the exact charged
accelerating BTZ black hole solutions are obtained in Ref. \cite%
{EslamPanahArXiv}.

It is interesting to find (un)charged accelerating black hole solutions in
modified theories of gravity, such as $F(R)$ gravity. Generally, the field
equations of $F(R)$ theory coupled to a matter field is complicated, and
hence it is not easy to find exact analytical accelerating black hole
solutions. In this regard, by considering $F(R)$ gravity, the charged
accelerating black hole solutions in four-dimensional spacetime had been
obtained in Ref. \cite{ZhangMann2019}. However, there is no
three-dimensional (un)charged accelerating black hole solution in the
context of $F(R)$ gravity. For this purpose, by coupling $F(R)$ gravity with
CIM, we will find (un)charged accelerating black hole solutions in this
theory known as $F(R)$-CIM gravity.

\section{Uncharged Solutions}

Here, we consider $F(R)$ gravity to extract accelerating black hole
solutions in three-dimensional spacetime. We consider the action of $F(R)$
gravity in the following form 
\begin{equation}
\mathcal{I}=\frac{1}{16\pi }\int_{\partial \mathcal{M}}d^{3}x\sqrt{-g}F(R),
\label{actionF(R)a}
\end{equation}%
where $F(R)=R+f\left( R\right) $, which $R$ is scalar curvature (Ricci
scalar), and $f\left( R\right) $ is an arbitrary function of scalar
curvature $R$. In the above action, we considered $G=1$.

Varying the action (\ref{actionF(R)a}) with respect to the gravitational
field $g_{\mu \nu }$ leads to 
\begin{equation}
R_{\mu \nu }\left( 1+f_{R}\right) -\frac{g_{\mu \nu }F(R)}{2}+\left( g_{\mu
\nu }\nabla ^{2}-\nabla _{\mu }\nabla _{\nu }\right) f_{R}=0,
\label{EqF(R)1a}
\end{equation}%
where $f_{R}=\frac{df(R)}{dR}$.

We consider the following three-dimensional $C-$metric to extract the
accelerating black hole solutions as it was introduced in Refs. \cite%
{BTZAcII,BTZAcIIb} 
\begin{equation}
ds^{2}=\frac{1}{\mathcal{K}^{2}\left( r,\theta \right) }\left[ -g(r,\theta
)dt^{2}+\frac{dr^{2}}{g(r,\theta )}+r^{2}d\theta ^{2}\right] ,
\label{Metric}
\end{equation}%
where $\mathcal{K}\left( r,\theta \right) =\alpha r\cosh \left( \sqrt{m-k}%
\theta \right) -1$, which is called the conformal factor. Also, $k$ is a
topological constant that can be $\pm 1$ or $0$. In addition, $g(r,\theta )$
is a metric function, which we have to find it. The coordinates of $t$ and $%
r $ in the metric, respectively, are in the ranges $-\infty <t<\infty $ and $%
0\leq r<\infty $. Due to the lack of translational symmetry in $\theta $ (in
the presence of the conformal factor), we seem to have no reason to restrict
it to be in $\left[ -\pi ,\pi \right] $. So, we consider $\theta $ in the
range $-\pi \leq \theta \leq \pi $ (more details are given in Ref. \cite%
{BTZAcII,BTZAcIIb}).

It is notable that in three-dimensional spacetime, we may find two methods
for understanding if our metric is under acceleration or not. These methods
are:

(i) a domain wall (which is studied by Arenas-Henriquez et al. in Ref. \cite%
{BTZAcIII}).

(ii) by calculating the proper acceleration of spacetime (which is discussed
in Refs. \cite{BTZAcI,BTZAcII,BTZAcIIb}).

Here, we consider the method (ii) and calculate the proper acceleration of
the mentioned spacetime (Eq. (\ref{Metric})). Considering a static observer
in spacetime as 
\begin{equation}
x^{\mu }\left( \lambda \right) =\left( \lambda \frac{\mathcal{K}}{\sqrt{%
g(r,\theta )}},r,\theta \right) ,  \label{coordinate}
\end{equation}%
where $\lambda $ is the proper time. Also, proper velocity is defined $%
u^{\mu }=\frac{dx^{\mu }}{d\lambda }$. Using Eq. (\ref{coordinate}) and the
definition of the proper velocity, we can get $u^{\mu} $ as 
\begin{equation}
u^{\mu }\left( \lambda \right) =\left( \frac{\mathcal{K}}{\sqrt{g(r,\theta )}%
},0,0\right) .
\end{equation}

We will calculate the proper acceleration after extracting the metric
function ($g(r,\theta )$).

In the following, we want to obtain the solutions for the constant scalar
curvature (i.e., $R=R_{0}=constant$) in three-dimensional spacetime. The
trace of Eq. (\ref{EqF(R)1a}) yields 
\begin{equation}
R_{0}\left( 1+f_{R_{0}}\right) -\frac{3}{2}\left( R_{0}+f(R_{0})\right) =0,
\end{equation}%
where $f_{R_{0}}=$ $f_{R_{\left\vert _{R=R_{0}}\right. }}$, and the solution
for $R_{0}$ gives 
\begin{equation}
R_{0}=\frac{3f(R_{0})}{2f_{R_{0}}-1}\equiv 6\Lambda <0.  \label{R0}
\end{equation}

Since it was shown that in three-dimensional dS spacetime, classical black
holes do not exist \cite{Emparan}, so we consider the negative value for $%
R_{0}$ (or the cosmological constant) in this work.

Substituting the equation (\ref{R0}) into Eq. (\ref{EqF(R)1a}), we can
obtain the equations of motion in $F(R)$ gravity in the following form 
\begin{equation}
R_{\mu \nu }\left( 1+f_{R_{0}}\right) -\frac{g_{\mu \nu }R_{0}\left(
1+f_{R_{0}}\right) }{3}=0.  \label{F(R)TraceI}
\end{equation}

Considering the metric (\ref{Metric}), and the field equation of $F(R)$
gravity (\ref{F(R)TraceI}), one can write the following field equations

\begin{eqnarray}
eq_{tt} &=&eq_{rr}=\left( \mathcal{K}+1\right) ^{2}\left( rg^{\prime \prime
}-2g^{\prime }+\frac{2g}{r}\right) +rg^{\prime \prime }-2\left( \mathcal{K}%
+1\right) \left( rg^{\prime \prime }-\frac{g^{\prime }}{2}-\frac{g}{r}%
\right)   \notag \\
&&+g^{\prime }+\frac{2rR_{0}}{3},  \label{1a} \\
&&  \notag \\
eq_{\theta \theta } &=&g^{\prime }-\frac{\left( \mathcal{K}+1\right) \left(
rg^{\prime }-2g\right) }{r}+\frac{rR_{0}}{3},  \label{2a}
\end{eqnarray}%
where $eq_{tt}$, $eq_{rr}$ and $eq_{\theta \theta }$, respectively, are
related to components of $tt$, $rr$ and $\theta \theta $ of Eq. (\ref%
{F(R)TraceI}). Also, in the above equations $\mathcal{K}=\mathcal{K}\left(
r,\theta \right) $, $g=g\left( r,\theta \right) $, $g^{\prime }=\frac{%
\partial g\left( r,\theta \right) }{\partial r}$, and $g^{\prime \prime }=%
\frac{\partial ^{2}g\left( r,\theta \right) }{\partial r^{2}}$. Now we are
in a position to obtain an exact solution from Eqs. (\ref{1a}) and (\ref{2a}%
). After some calculations, one can get the following metric function 
\begin{equation}
g(r,\theta )=C\left( \theta \right) \mathcal{K}^{2}-\frac{\left( 2\mathcal{K}%
+1\right) r^{2}R_{0}}{6\left( \mathcal{K}+1\right) ^{2}},  \label{Sol1}
\end{equation}%
where the above solution (\ref{Sol1}) satisfies all components of the field
equation (\ref{F(R)TraceI}). Also, $C\left( \theta \right) $ is an
integration constant-dependent and depends on $\theta $. It is necessary to
mention that the solutions in any modified theories of gravity must reduce
to the famous solutions in Einstein's gravity. The obtained solution (\ref%
{Sol1}) does not reduce to the famous BTZ black hole solutions when $\alpha
=0$. Using a choice for $C\left( \theta \right) $, we can remove this
problem. This choice is 
\begin{equation}
C\left( \theta \right) =k-m-\frac{R_{0}}{6\alpha ^{2}\cosh ^{2}\left( \sqrt{%
m-k}\theta \right) },  \label{C1a}
\end{equation}%
and by replacing Eq. (\ref{C1a}) into the solution (\ref{Sol1}), we obtain
accelerating black holes in $F(R)$ gravity in the following form 
\begin{equation}
g(r,\theta )=\left( k-m\right) \mathcal{K}^{2}-\frac{r^{2}R_{0}}{6},
\label{g(r)F(R)a}
\end{equation}%
where in the absence of the acceleration parameter (i.e., $\alpha =0$), the
solution (\ref{g(r)F(R)a}) reduces to the BTZ black hole in $F(R)$, as we
expected. It is notable that for recovering the famous static BTZ black hole
and to respect the signatures of spacetime (\ref{Metric}), we have to
restrict $m>k$.

Now we are in a position to get the proper acceleration. After some
calculations, we can obtain the proper acceleration (which is defined as $%
a^{\mu }=u^{\nu }\nabla _{\nu }u^{\mu }$) for the observer in the distance
as 
\begin{equation}
a^{\mu }a_{\mu }=\frac{R_{0}^{2}\left[ 6\left( k-m\right) \alpha ^{2}-R_{0}%
\right] }{\left( 6\alpha ^{2}\left( k-m\right) \cosh ^{2}\left( \sqrt{m-k}%
\theta \right) -R_{0}\right) ^{2}},
\end{equation}%
in which indicates, $\alpha $ is proportional (but not equal) to the
magnitude $|a|$ of the proper acceleration (see Refs. \cite%
{BTZAcI,BTZAcII,BTZAcIIb} for more details). So, we call $\alpha $ as an
acceleration parameter.

Considering the solution (\ref{g(r)F(R)a}), we want to find the Kretschmann
scalar of the spacetime (\ref{Metric}) which leads to 
\begin{equation}
R_{\alpha \beta \gamma \delta }R^{\alpha \beta \gamma \delta }=\frac{%
R_{0}^{2}}{3},
\end{equation}%
where indicates that this scalar does not diverge at $r=0$. Indeed, a
curvature singularity does not exist at $r=0$, but there is a conical
singularity for this spacetime.

In the following, we find the real roots of the metric function (\ref%
{g(r)F(R)a}). In this regard, we solve the metric function (\ref{g(r)F(R)a})
and get two roots in the following forms 
\begin{equation}
r_{root_{\pm }}=\frac{2\left[ 3\alpha \left( m-k\right) \left( \mathcal{B}%
^{2}+1\right) \pm \mathcal{B}\sqrt{6R_{0}\left( k-m\right) }\right] \mathcal{%
B}}{2R_{0}\mathcal{B}^{2}+3\alpha ^{2}\left( m-k\right) \left( \mathcal{B}%
^{4}+2\mathcal{B}^{2}+1\right) },
\end{equation}%
where $\mathcal{B}=\exp \left( \sqrt{m-k}\theta \right)$. To have the real
root for the metric function (\ref{g(r)F(R)a}), we have to respect $R_{0}<0$%
, since $m>k$.

\begin{figure}[tbph]
\centering
\includegraphics[width=0.4\linewidth]{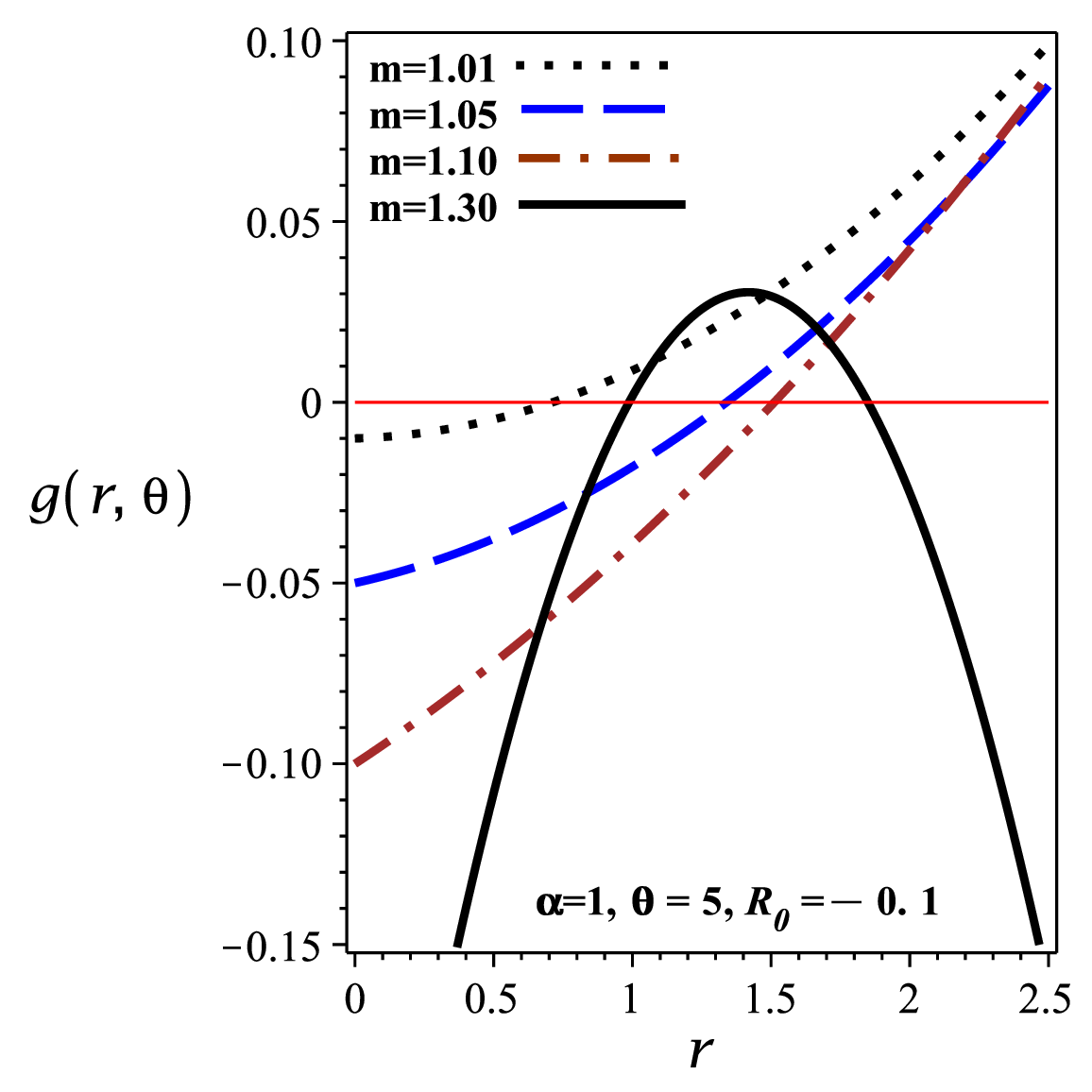} \includegraphics[width=0.4%
\linewidth]{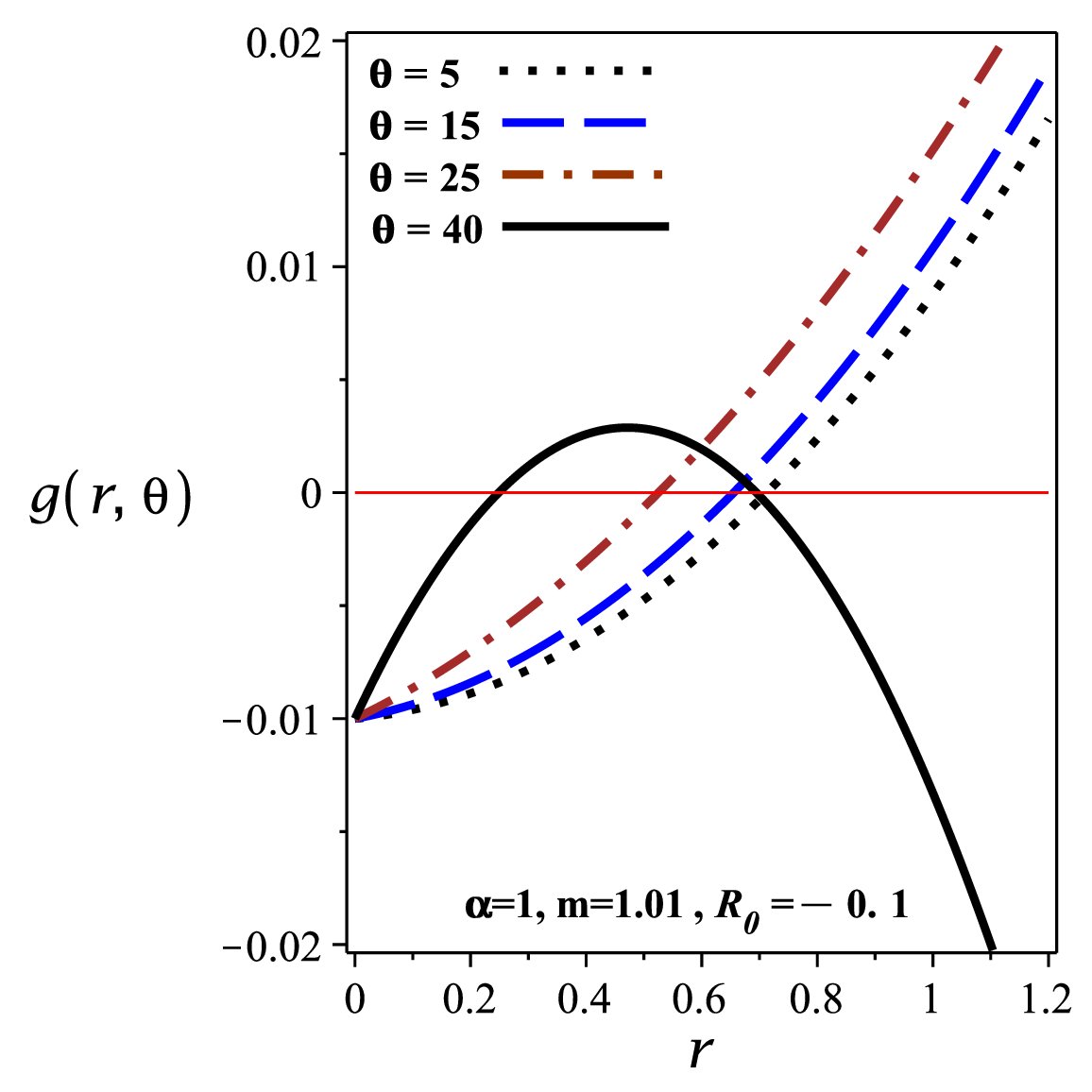} \includegraphics[width=0.4\linewidth]{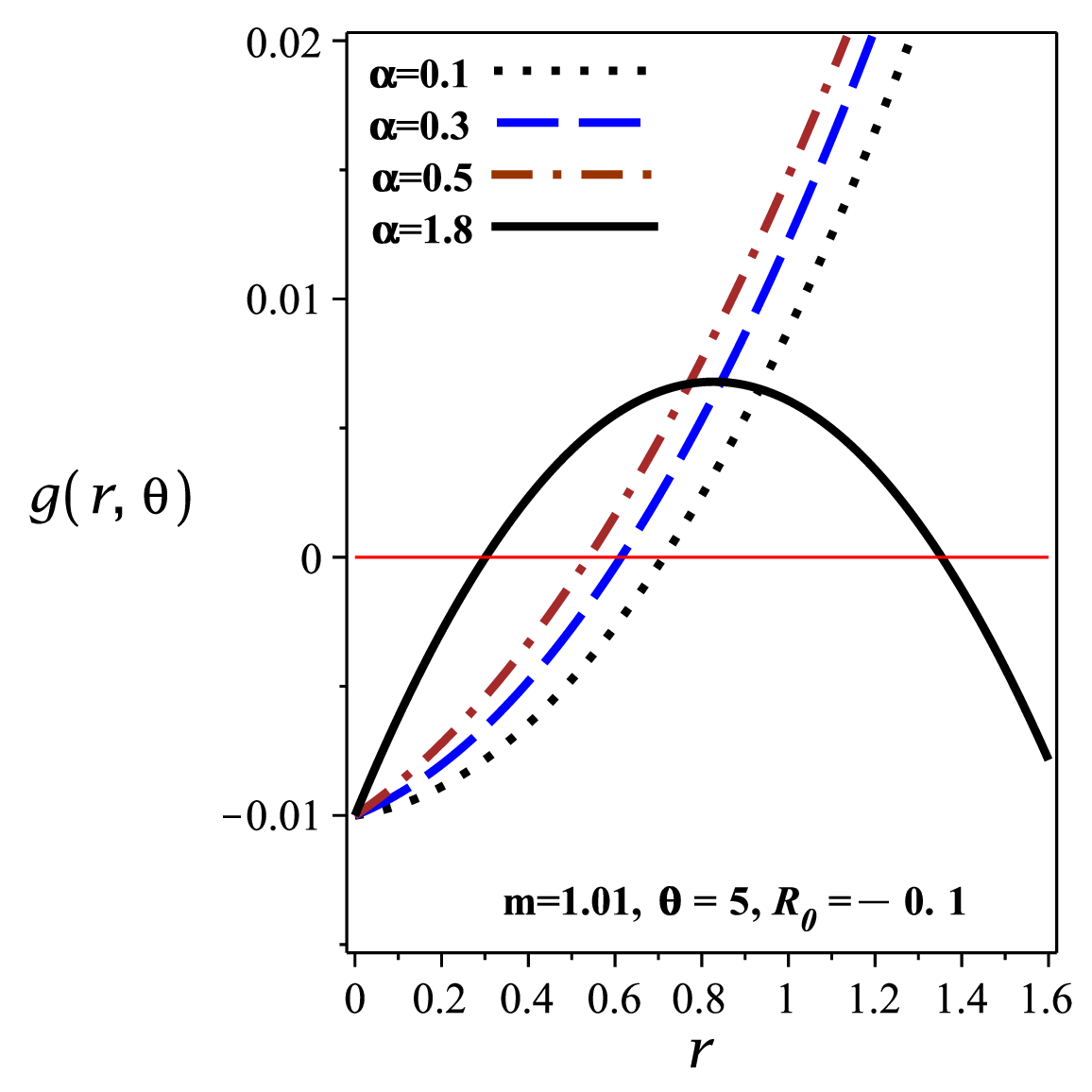} %
\includegraphics[width=0.4\linewidth]{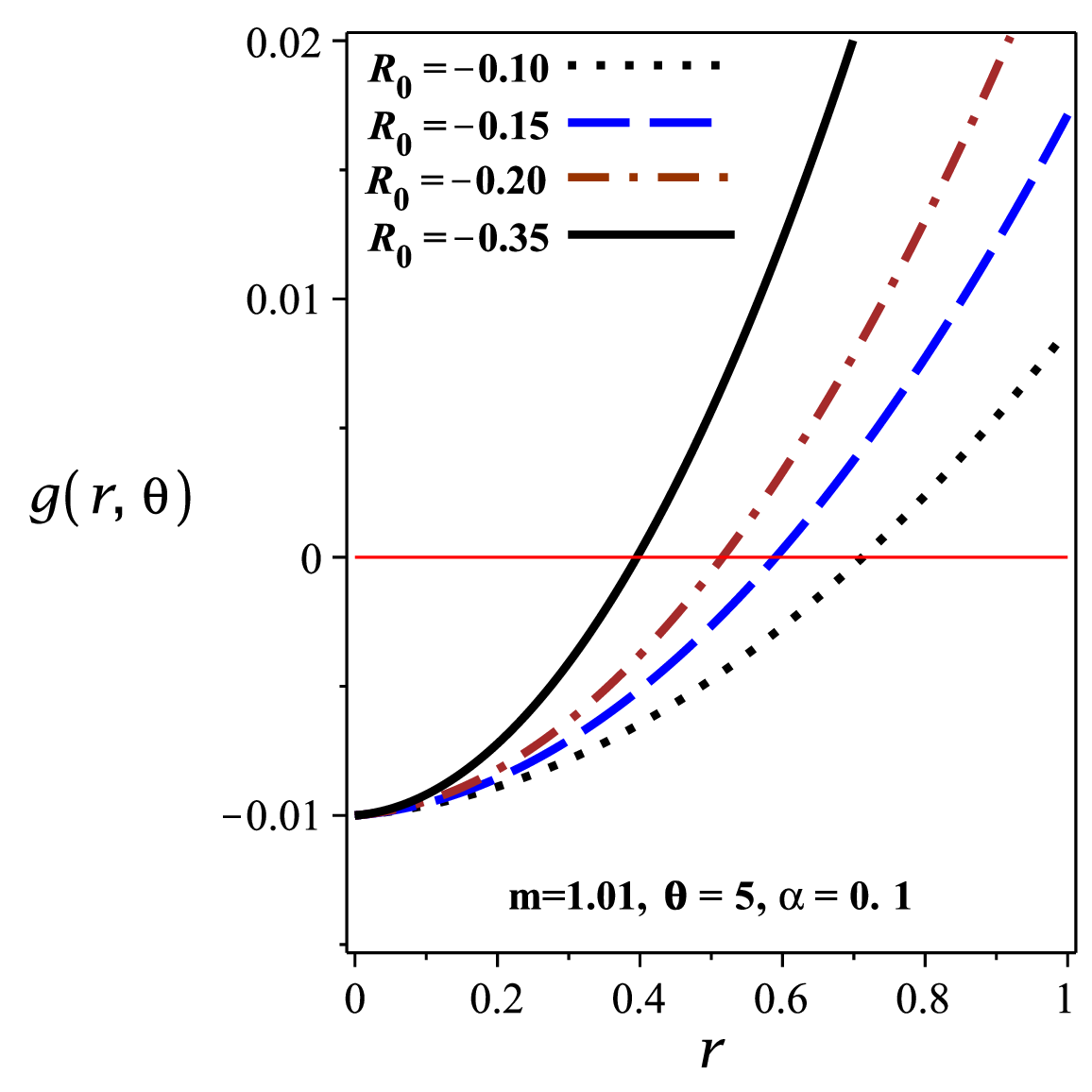}
\caption{The metric function $g(r,\protect\theta )$ versus $r$ for $k=1$ and
different values of the parameters.}
\label{Fig1a}
\end{figure}

Here, we plot $g(r,\theta )$\ versus $r$\ in Fig. \ref{Fig1a}. As shown in
Fig. \ref{Fig1a}, there is at least a root. In order to investigate the
effects of various parameters on the roots, we plotted four diagrams in Fig. %
\ref{Fig1a}. Each diagram indicates that each parameter how affects the
roots.

Our results indicate that with increasing mass, the radii of the
accelerating BTZ AdS black holes increase (see the dotted, dashed, and
dotted-dashed lines in the up-left panel in Fig. \ref{Fig1a}). However,
there is a critical value for mass ($m_{critical}$) which by increasing it
more than a critical value, i.e., $m>m_{critical}$, these black holes
include two roots. The small root is related to the event horizon, and the
large root belongs to the cosmological horizon. The effect of $\theta $ in
the up-right panel in Fig. \ref{Fig1a} reveals that by increasing $\theta $,
the radius of the black hole decreases. In addition, there is a critical
value for $\theta $ ($\theta _{critical}$) which by increasing this quantity
more than a critical value, i.e., $\theta >\theta _{critical}$, these black
holes include two horizons (event and cosmological horizons). There is the
same behavior for the acceleration parameter. Indeed, by increasing $\alpha $%
, the radius of the black hole decreases. Also, there is a critical value
for $\alpha>\alpha _{critical}$. In this case, the black holes have two
roots. The small root is related to the event horizon, and the large root
belongs to the cosmological horizon (see the down-left panel in Fig. \ref%
{Fig1a}). Also, we encounter small black holes when the absolute value of
the Ricci scalar $\left\vert R_{0}\right\vert $\ increase (see the
down-right panels in Fig. \ref{Fig1a}).

Considering the obtained metric function, the asymptotical behavior of the
mentioned spacetime depends on the parameters of this theory, which include $%
\alpha $, $k$, $m$, and $R_{0}$. As a result, the obtained spacetime is not
exactly asymptotically AdS due to different parameters in the obtained
metric function.

Here, we study the effects of different parameters on the temperature of
these black holes. For this purpose, we first have to get the black hole's
temperature. Using the Hawking temperature in the following form 
\begin{equation}
T_{H}=\frac{\kappa }{2\pi },  \label{THa}
\end{equation}%
where $\kappa $\ is the surface gravity in the form $\kappa =\sqrt{-\frac{1}{%
2}\left( \nabla _{\mu }\chi _{\nu }\right) \left( \nabla ^{\mu }\chi ^{\nu
}\right) }$. Also, $\chi =\partial _{t}$\ is the Killing vector. Considering
the metric (\ref{Metric}), we get the surface gravity as 
\begin{equation}
\kappa =\left. \frac{1}{2}\left( \frac{\partial g\left( r,\theta \right) }{%
\partial r}\right) \right\vert _{r=r_{+}}.  \label{kappaI}
\end{equation}

Using the above equation and the metric function (\ref{g(r)F(R)a}), we are
in a position to get the Hawking temperature. To obtain the temperature of
these black holes, we have to express the mass ($m$) in terms of the radius
of event horizon $r_{+}$, the accelerating parameter $\alpha $, and $R_{0}$\
in the following form 
\begin{equation}
m=k-\frac{r_{+}^{2}R_{0}}{6\left( \alpha r_{+}-1\right) ^{2}},  \label{mha}
\end{equation}%
which we extract the above equation by considering $g\left( r,\theta \right)
=0$. Notably, to get the exact form (\ref{mha}), we suppose $\theta =0$\ in
the metric function (\ref{g(r)F(R)a}).

Substituting the mass (\ref{mha}) within the equations (\ref{kappaI}) and
then (\ref{THa}), one can calculate the Hawking temperature 
\begin{equation}
T_{H}=\frac{r_{+}R_{0}}{12\pi \alpha \left( \alpha r_{+}-1\right) }.
\label{TbtzFRa}
\end{equation}

In the context of black holes, it is argued that the root of temperature ($%
T=0$) represents a borderline between physical ($T>0$) and non-physical ($%
T<0 $) black holes \cite{Temperature}. To study the behavior of temperature
of these black holes, we evaluate their roots. There is no root for the
temperature. On the other hand, to have a positive temperature, we must
respect $\alpha r_{+}<1$ because $R_{0}<0$. By considering $\alpha r_{+}<1$,
the temperature increases by increasing (decreasing) the value of $%
\left\vert R_{0}\right\vert $ ($\alpha $).

\section{Charged Solutions}

In order to extract charged accelerating black hole solutions in
three-dimensional spacetime, we add the CIM field as a source (which leads
to traceless energy-momentum tensor) to $F(R)$ gravity. The action of $F(R)$
gravity coupled to the PMI field is given \cite{HendiES2014} 
\begin{equation}
\mathcal{I}=\frac{1}{16\pi }\int_{\partial \mathcal{M}}d^{3}x\sqrt{-g}\left[
F(R)+L\left( \mathcal{F}\right) \right] ,  \label{actionF(R)}
\end{equation}%
where $L\left( \mathcal{F}\right) =\left( -\mathcal{F}\right) ^{s}$, and for
CIM filed $s=\frac{3}{4}$ (i.e., $L\left( \mathcal{F}\right) =\left( -%
\mathcal{F}\right) ^{\frac{3}{4}}$). $\mathcal{F}=$\ $F_{\mu \nu }F^{\mu \nu
}$ is the Maxwell invariant ($F_{\mu \nu }=\partial _{\mu }A_{\nu }-\partial
_{\nu }A_{\mu }$ is the electromagnetic tensor field, and $A_{\mu }$ is the
gauge potential). Varying the action (\ref{actionF(R)}) with respect to the
gravitational field $g_{\mu \nu }$, and the gauge field $A_{\mu }$, lead to 
\begin{eqnarray}
R_{\mu \nu }\left( 1+f_{R}\right) -\frac{g_{\mu \nu }F(R)}{2}+\left( g_{\mu
\nu }\nabla ^{2}-\nabla _{\mu }\nabla _{\nu }\right) f_{R} &=&8\pi \mathrm{T}%
_{\mu \nu },  \label{EqF(R)1} \\
&&  \notag \\
\partial _{\mu }\left( \sqrt{-g}\mathcal{F}^{-\frac{1}{4}}F^{\mu \nu
}\right) &=&0,  \label{Eqch2}
\end{eqnarray}%
where the energy-momentum tensor ($\mathrm{T}_{\mu \nu }$) in the presence
of CIM for three-dimensional spacetime can be written 
\begin{equation}
\mathrm{T}_{\mu \nu }=\frac{3}{2}F_{\mu \rho }F_{\nu }^{\rho }\left( -%
\mathcal{F}\right) ^{-\frac{1}{4}}+\frac{1}{2}g_{\mu \nu }\left( -\mathcal{F}%
\right) ^{\frac{3}{4}}.  \label{Tmunu}
\end{equation}

We obtain the equations of motion in $F(R)$-CIM gravity by replacing the
equation (\ref{R0}) into Eq. (\ref{EqF(R)1}), which leads to 
\begin{equation}
R_{\mu \nu }\left( 1+f_{R_{0}}\right) -\frac{g_{\mu \nu }R_{0}\left(
1+f_{R_{0}}\right) }{3}=8\pi \mathrm{T}_{\mu \nu }.  \label{F(R)Trace}
\end{equation}

Notably, the assumption of a traceless energy-momentum tensor (a result of
CIM) is essential for deriving exact black hole solutions in $F(R)$ gravity
coupled to the matter field in metric formalism.

According to this fact that we look for black hole solutions with a radial
electric field, so the gauge potential is given by $A_{\mu }=h(r)\delta
_{\mu }^{0}$ (i.e., $A_{\mu }=\left[ h(r),0,0\right] $). Using the
introduced metric in the equation (\ref{Metric}) and the mentioned gauge
potential, one can show that Eq. (\ref{Eqch2}) reduces to 
\begin{equation}
\mathcal{K}^{4}\left( r,\theta \right) \left( rh^{\prime \prime
}(r)+2h^{\prime }(r)\right) =0,  \label{Eqh(r)}
\end{equation}%
solving Eq. (\ref{Eqh(r)}), one can find 
\begin{equation}
h(r)=-\frac{q}{r},  \label{h(r)}
\end{equation}%
where $q$ is an integration constant related to the electric charge. Our
calculation for obtaining the electric field leads to $F_{tr}=-F_{rt}=%
\partial _{t}A_{r}-\partial _{r}A_{t}=\frac{q}{r^{2}}$, where it shows that
the electric field in three-dimensional spacetime is the inverse square of $%
r $ (similar to four-dimensional spacetime). It is notable that we restrict
ourselves to $\alpha r\cosh \left( \sqrt{m-k}\theta \right) \neq 1$ to
obtain the equation (\ref{h(r)}).

Considering the metric (\ref{Metric}), the electric field, and the field
equation of $F(R)$-CIM gravity (\ref{F(R)Trace}), one can write the
following field equations

\begin{eqnarray}
eq_{tt} &=&eq_{rr}=\frac{\left( 2q^{2}\right) ^{\frac{3}{4}}\mathcal{K}^{3}}{%
2\left( 1+f_{R_{0}}\right) r^{2}}-\frac{\left( \mathcal{K}+1\right)
^{2}\left( r^{2}g^{\prime \prime }-2rg^{\prime }+2g\right) }{r}+\frac{%
2\left( \mathcal{K}+1\right) \left( r^{2}g^{\prime \prime }-\frac{rg^{\prime
}}{2}-g\right) }{r}  \notag \\
&&-rg^{\prime \prime }-g^{\prime }-\frac{2rR_{0}}{3},  \label{eq11} \\
&&  \notag \\
eq_{\theta \theta } &=&\frac{\left( 2q^{2}\right) ^{\frac{3}{4}}\mathcal{K}%
^{3}}{2\left( 1+f_{R_{0}}\right) r^{2}}+g^{\prime }+\frac{rR_{0}}{3}-\frac{%
\left( \mathcal{K}+1\right) \left( rg^{\prime }-2g\right) }{r},  \label{eq33}
\end{eqnarray}%
and we get the metric function as 
\begin{equation}
g(r,\theta )=C\left( \theta \right) \mathcal{K}^{2}-\frac{\left( 2\mathcal{K}%
+1\right) r^{2}R_{0}}{6\left( \mathcal{K}+1\right) ^{2}}-\frac{\left(
2q^{2}\right) ^{\frac{3}{4}}\mathcal{K}^{2}}{2r\left( 1+f_{R_{0}}\right) },
\label{f(r)Uch}
\end{equation}%
where the solution (\ref{f(r)Uch}) satisfies all components of the field
equation (\ref{F(R)Trace}). In order to recover the famous solutions in
Einstein's gravity, we consider 
\begin{equation}
C\left( \theta \right) =k-m-\frac{R_{0}}{6\alpha ^{2}\cosh ^{2}\left( \sqrt{%
m-k}\theta \right) }-\frac{\left( 2q^{2}\right) ^{\frac{3}{4}}\alpha \cosh
\left( \sqrt{m-k}\theta \right) }{4\left( 1+f_{R_{0}}\right) },  \label{C1}
\end{equation}%
and by replacing Eq. (\ref{C1}) into the solution (\ref{f(r)Uch}), we obtain
charged accelerating black hole solutions in $F(R)$ gravity 
\begin{equation}
g(r,\theta )=\left( k-m\right) \mathcal{K}^{2}-\frac{\left( 2q^{2}\right) ^{%
\frac{3}{4}}\mathcal{K}^{2}\left( \mathcal{K}+3\right) }{4\left(
1+f_{R_{0}}\right) r}-\frac{r^{2}R_{0}}{6},  \label{g(r)F(R)}
\end{equation}%
where we restrict ourselves to $f_{R_{0}}\neq -1$. Indeed, to have physical
solutions, we should restrict ourselves to $f_{R_{0}}\neq -1$. In the
absence of the acceleration parameter (i.e., $\alpha =0$), the solution (\ref%
{g(r)F(R)}) reduces to the BTZ black hole in $F(R)$-CIM gravity \cite%
{HendiES2014} in the following form, as we expected 
\begin{equation}
g(r) =k-m-\frac{\left( 2q^{2}\right) ^{\frac{3}{4}}}{2\left(
1+f_{R_{0}}\right) r}-\frac{r^{2}R_{0}}{6}.  \label{MBTZacc}
\end{equation}

Considering the introduced spacetime in Eq. (\ref{Metric}), with the metric
function (\ref{g(r)F(R)}), we obtain the Kretschmann scalar as $R_{\alpha
\beta \gamma \delta }R^{\alpha \beta \gamma \delta }=\frac{R_{0}^{2}}{3}+%
\frac{3\sqrt{2}q^{3}\mathcal{K}^{6}}{\left( 1+f_{R_{0}}\right) r^{6}}$,
where $\underset{r\longrightarrow 0}{\lim }R_{\alpha \beta \gamma \delta
}R^{\alpha \beta \gamma \delta }\longrightarrow \infty $, and indicates this
scalar diverge at $r=0$. So, there is a curvature singularity at $r=0$.

\begin{figure}[tbph]
\centering
\includegraphics[width=0.31\linewidth]{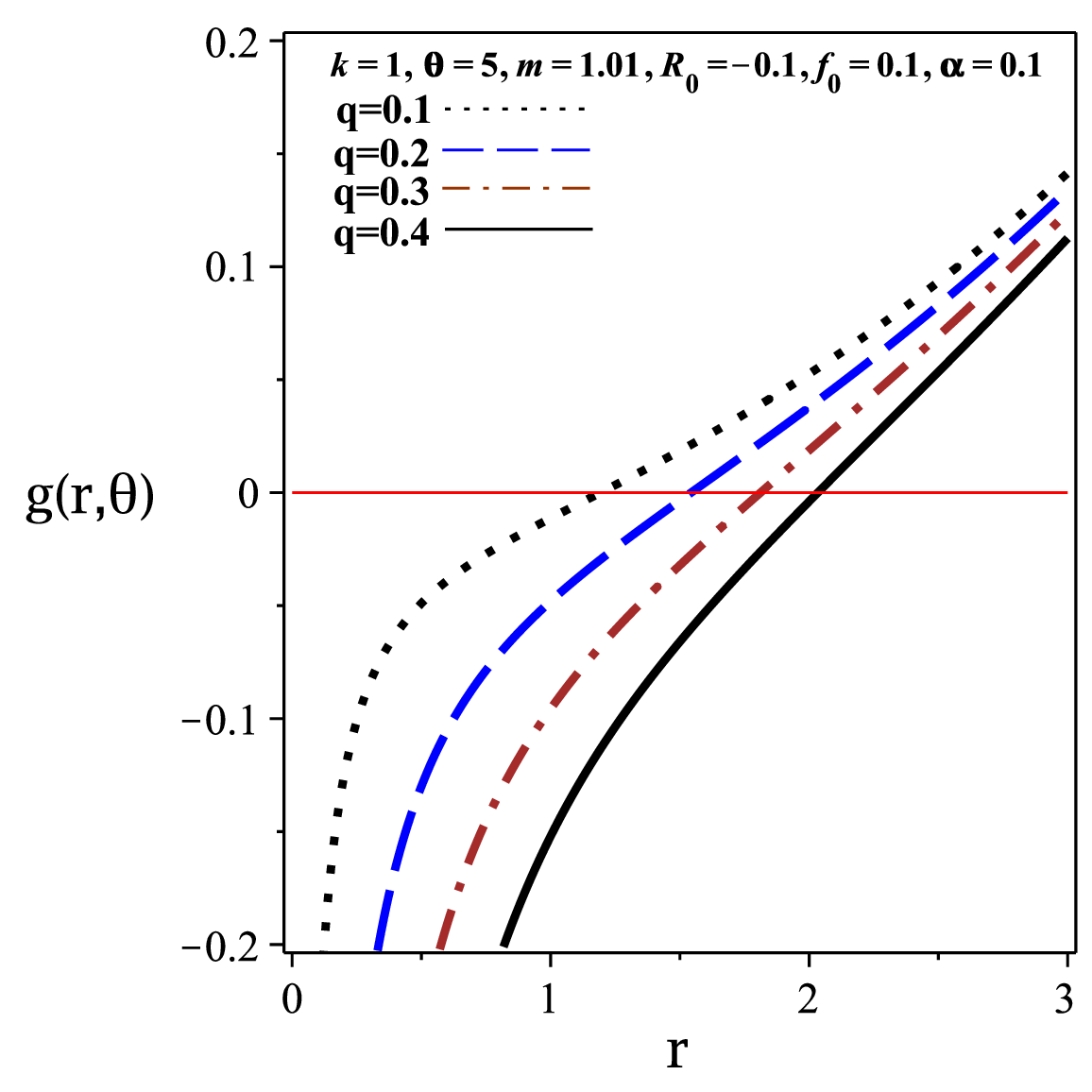} \includegraphics[width=0.31%
\linewidth]{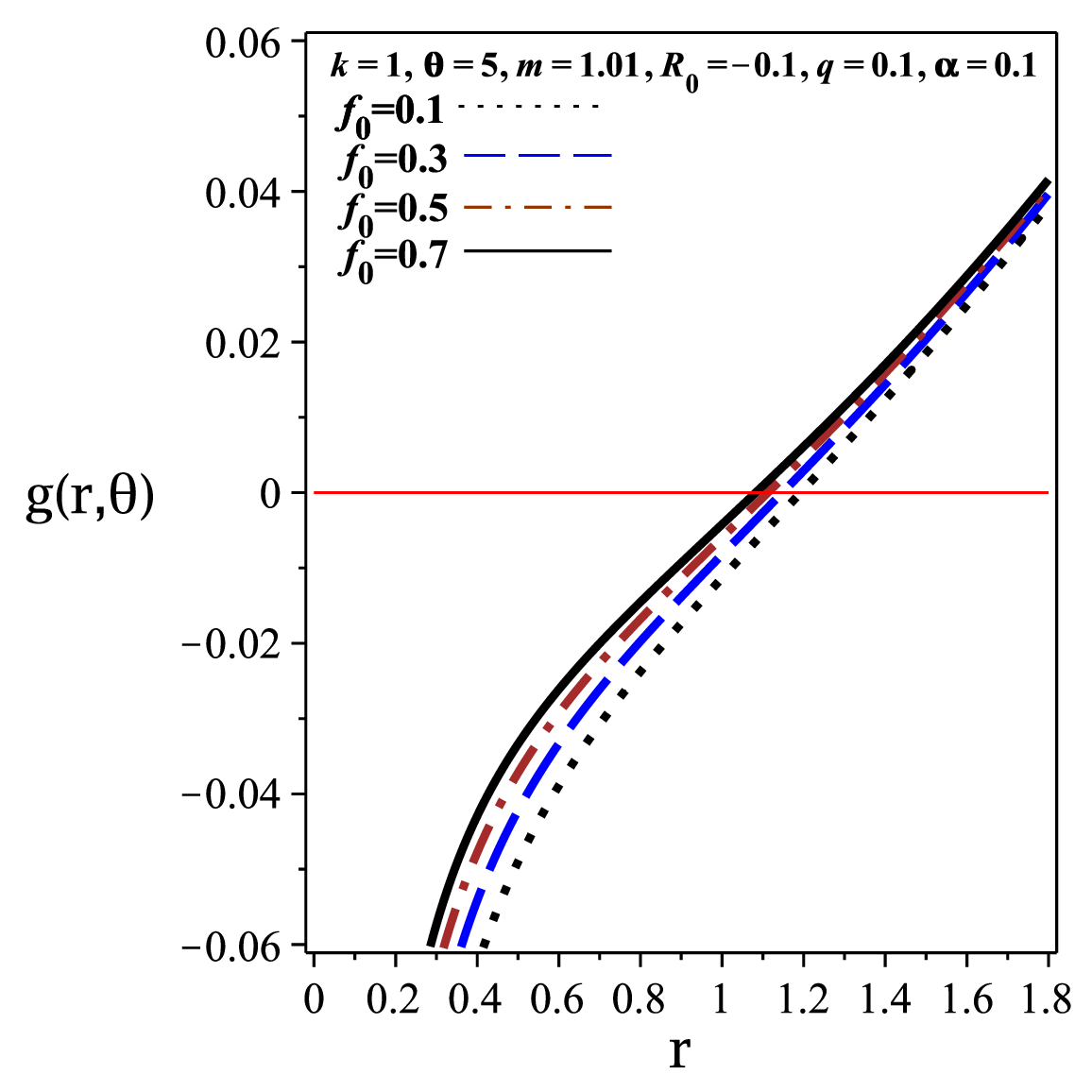} \includegraphics[width=0.31\linewidth]{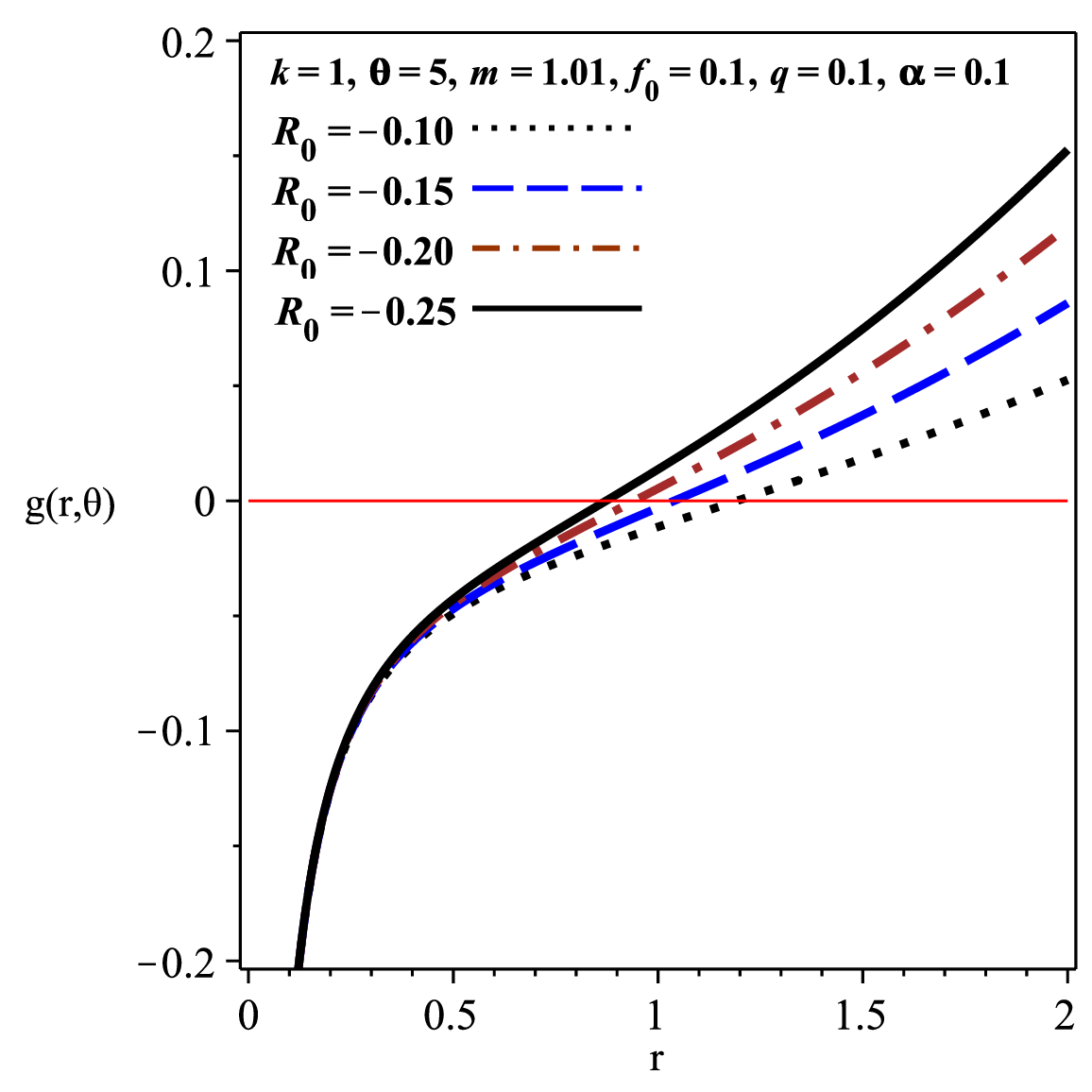} %
\includegraphics[width=0.31\linewidth]{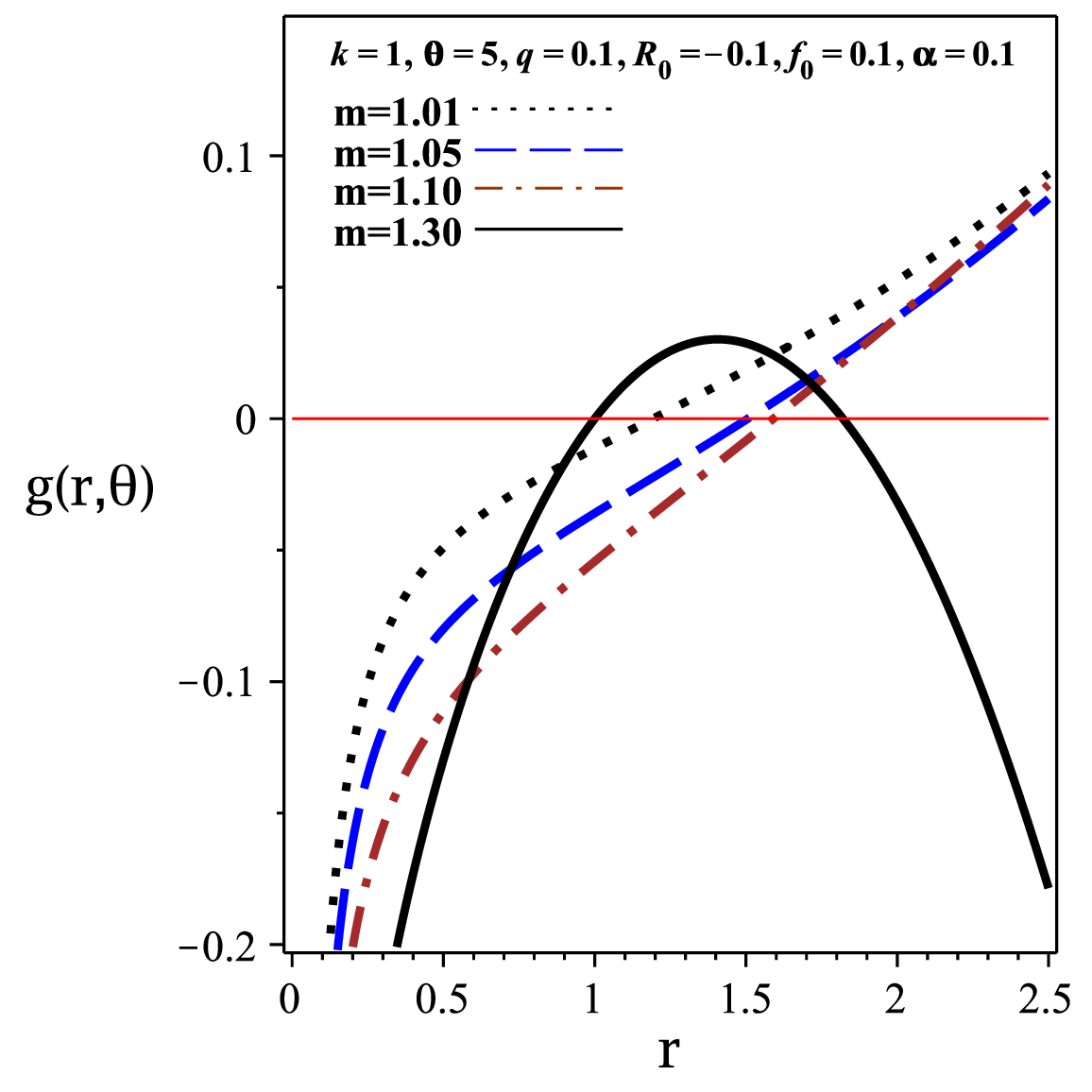} \includegraphics[width=0.31%
\linewidth]{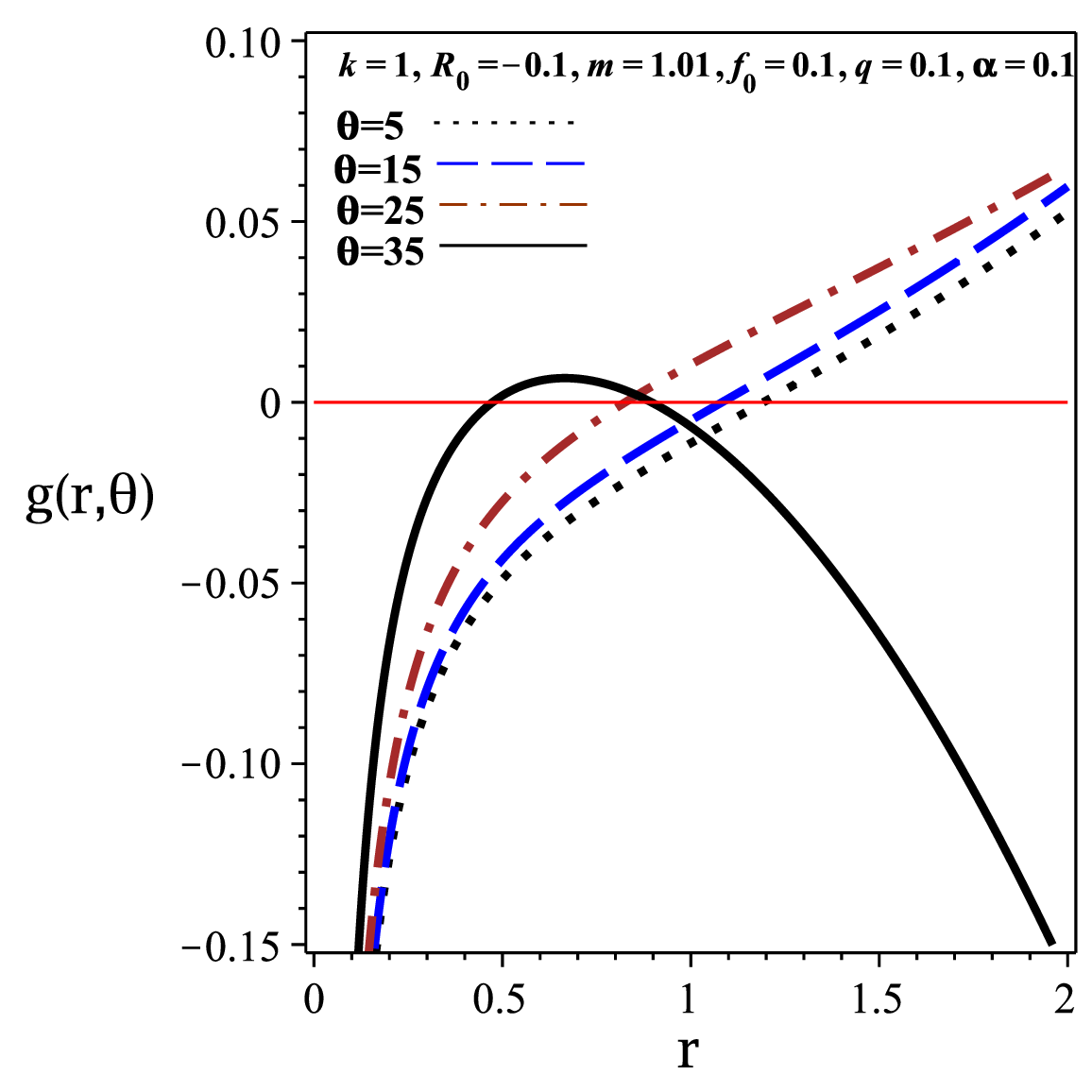} \includegraphics[width=0.31\linewidth]{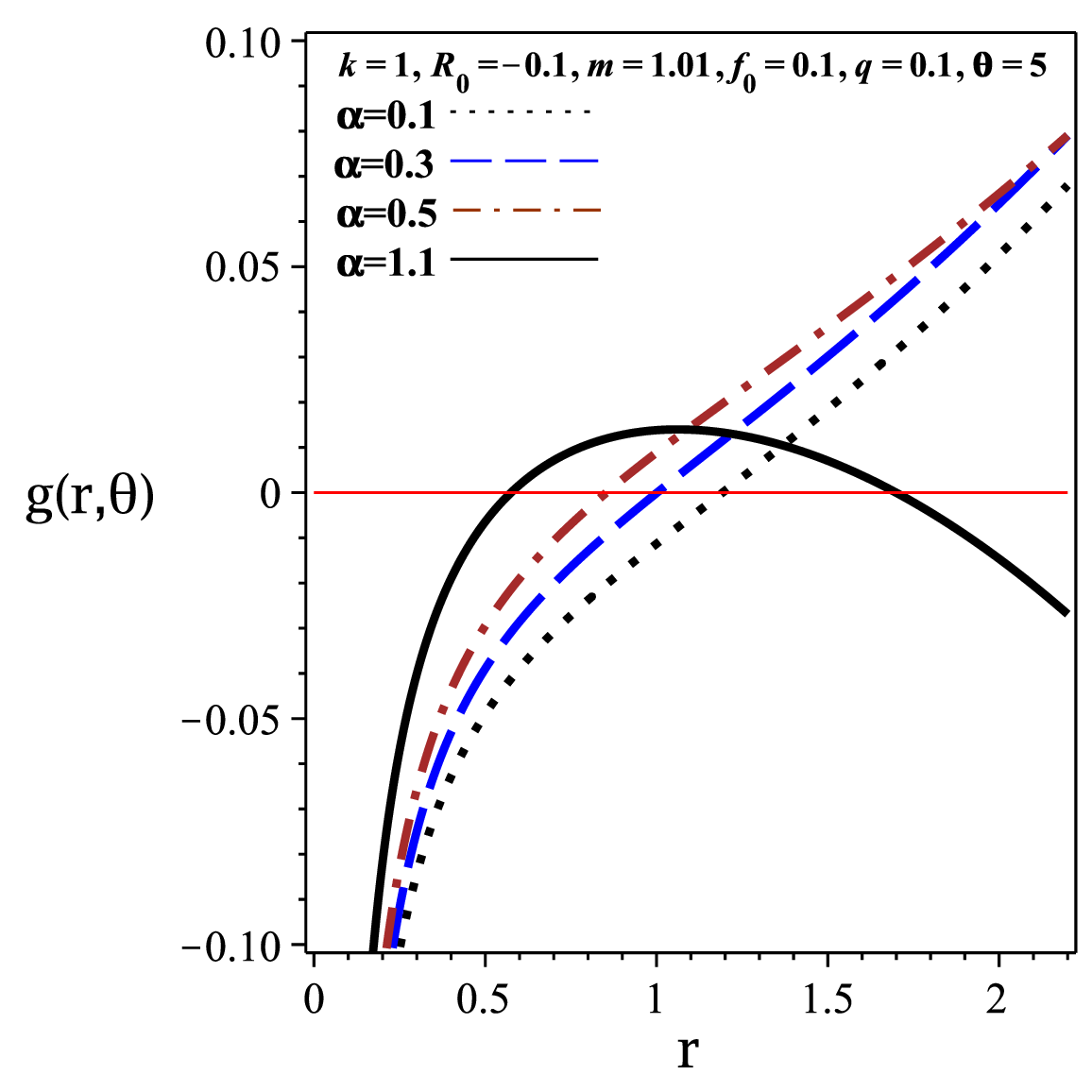}
\caption{The metric function $g(r,\protect\theta )$ versus $r$ for different
values of the parameters.}
\label{Fig1}
\end{figure}

In order to investigate the effects of various parameters on the roots of
the metric function (\ref{g(r)F(R)}), we plotted six diagrams in Fig. \ref%
{Fig1}. Each diagram indicates that each parameter how affects the roots.
The up-left panel in Fig. \ref{Fig1} shows that higher-charged accelerating
BTZ black holes have large radii. However, we encounter small black holes
when the parameter of $F(R)$ (i.e., $f_{0}$) and the absolute value of the
Ricci scalar $\left\vert R_{0}\right\vert $ increase (see up-middle and
up-right panels in Fig. \ref{Fig1}). Our results in the down panels of Fig. %
\ref{Fig1} show that there are the critical values for parameters of $m$, $%
\theta $, and $\alpha $. As shown in the down-left panel in Fig. \ref{Fig1},
massive charged accelerating black holes have large radii. However, by
increasing mass more than a critical value (i.e., $m>m_{critical}$), our
solutions include two roots. Indeed, The small root is related to the event
horizon, and the large root belongs to the cosmological horizon. We
encounter small black holes when $\theta $ and $\alpha $ increase. In other
words, the radius of the black hole decreases by increasing $\theta $ and $%
\alpha $. Moreover, there are two roots when $\theta $ and $\alpha $
increase more than a critical value (see continuous lines in the down-middle
and down-right panels in Fig. \ref{Fig1}).

Notably, the asymptotical behavior of the mentioned spacetime is dependent
on the parameters of this theory ($k$, $m$, $R_{0}$, $q$, and $f_{R_{0}}$).

Using the Hawking temperature in Eq. (\ref{THa}), and the metric function (%
\ref{g(r)F(R)}), we are in a position to get the Hawking temperature of
these black holes. We express the mass ($m$) in terms of the radius of the
event horizon $r_{+}$, the accelerating parameter $\alpha $, $R_{0}$ and the
charge $q$ in the following form 
\begin{equation}
m=k-\frac{r_{+}^{2}R_{0}}{6\left( \alpha r_{+}-1\right) ^{2}}-\frac{\left(
2q^{2}\right) ^{\frac{3}{4}}\left( \alpha r_{+}+2\right) }{4\left(
1+f_{R_{0}}\right) r_{+}},  \label{mh}
\end{equation}%
which we suppose $\theta =0$ in the metric function (\ref{g(r)F(R)}), to get
the exact form (\ref{mh}).

Substituting the mass (\ref{mh}) within the equation (\ref{THa}), one can
calculate the Hawking temperature 
\begin{equation}
T_{H}=\frac{r_{+}R_{0}}{12\pi \alpha \left( \alpha r_{+}-1\right) }+\frac{%
\left( 2q^{2}\right) ^{\frac{3}{4}}\left( \alpha r_{+}-1\right) ^{2}}{8\pi
\alpha \left( 1+f_{R_{0}}\right) r_{+}^{2}}.  \label{TbtzFR}
\end{equation}

To study the behavior of temperature of these black holes, we evaluate the
roots of it. Considering $T_{H}=0$ in Eq. (\ref{TbtzFR}), we get a root for
the temperature as 
\begin{equation}
r_{root}=\frac{1-\frac{2\left( 1+f_{R_{0}}\right) R_{0}}{\alpha \left(
6q\left( 1+f_{R_{0}}\right) R_{0}\mathcal{C}\right) ^{\frac{1}{3}}}+\frac{%
\left( 6q\left( 1+f_{R_{0}}\right) R_{0}\mathcal{C}\right) ^{\frac{1}{3}}}{%
3\alpha ^{2}\left( 2q^{2}\right) ^{\frac{3}{4}}}}{\alpha +\frac{2\left(
1+f_{R_{0}}\right) R_{0}}{3\alpha ^{2}\left( 2q^{2}\right) ^{\frac{3}{4}}}},
\end{equation}%
where $\mathcal{C}$ is given by 
\begin{equation}
\mathcal{C}=3\sqrt{2q\left[ \left( 2q^{2}\right) ^{\frac{3}{4}}\mathcal{C}%
_{1}+\frac{\sqrt{2}\left( 1+f_{R_{0}}\right) ^{2}R_{0}^{2}}{9}\right] }%
+\left( 2q^{2}\right) ^{\frac{3}{4}}\left( \frac{\left( 1+f_{R_{0}}\right)
R_{0}}{q}-\frac{3\sqrt{q}\alpha ^{3}}{2^{\frac{1}{4}}}\right) ,
\end{equation}%
also, $\mathcal{C}_{1}=\frac{\sqrt{2}\alpha ^{3}\left( 1+f_{R_{0}}\right)
R_{0}}{3}+\frac{q^{\frac{3}{2}}\alpha ^{6}}{2^{\frac{3}{4}}}$. It is clear
that the temperature and its root depend on $q$, $R_{0}$, $\alpha $ and $%
f_{R_{0}}$.

On the other hand, by solving $T_{H}$ versus $R_{0}$, we can get 
\begin{equation}
R_{0}=\frac{-3\left( 2q^{2}\right) ^{3/4}\left( \alpha r_{+}-1\right) ^{3}}{%
2\left( 1+f_{R_{0}}\right) r_{+}^{3}},
\end{equation}%
where $\left\vert f_{R_{0}}\right\vert <1$, because $F(R)$ gravity acts as a
modified theory, and so the values of $\left\vert f_{R_{0}}\right\vert $
cannot be more than $1$. In addition, by considering the AdS case (i.e., $%
R_{0}<0$), we must respect the condition $\alpha r_{+}>1$. Therefore, there
is a constraint on the acceleration parameter.

\begin{figure}[tbh]
\centering
\includegraphics[width=0.32\linewidth]{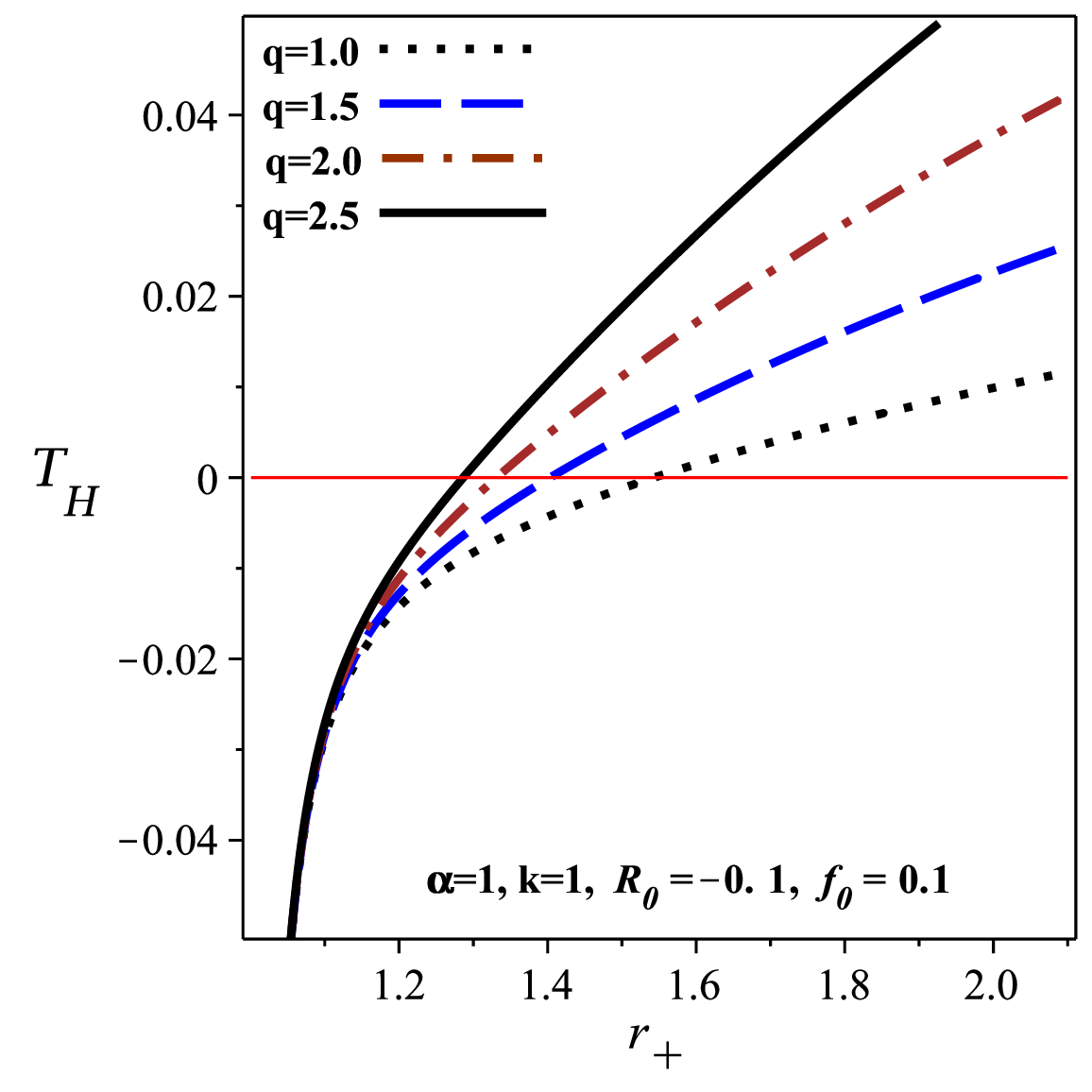} \includegraphics[width=0.32%
\linewidth]{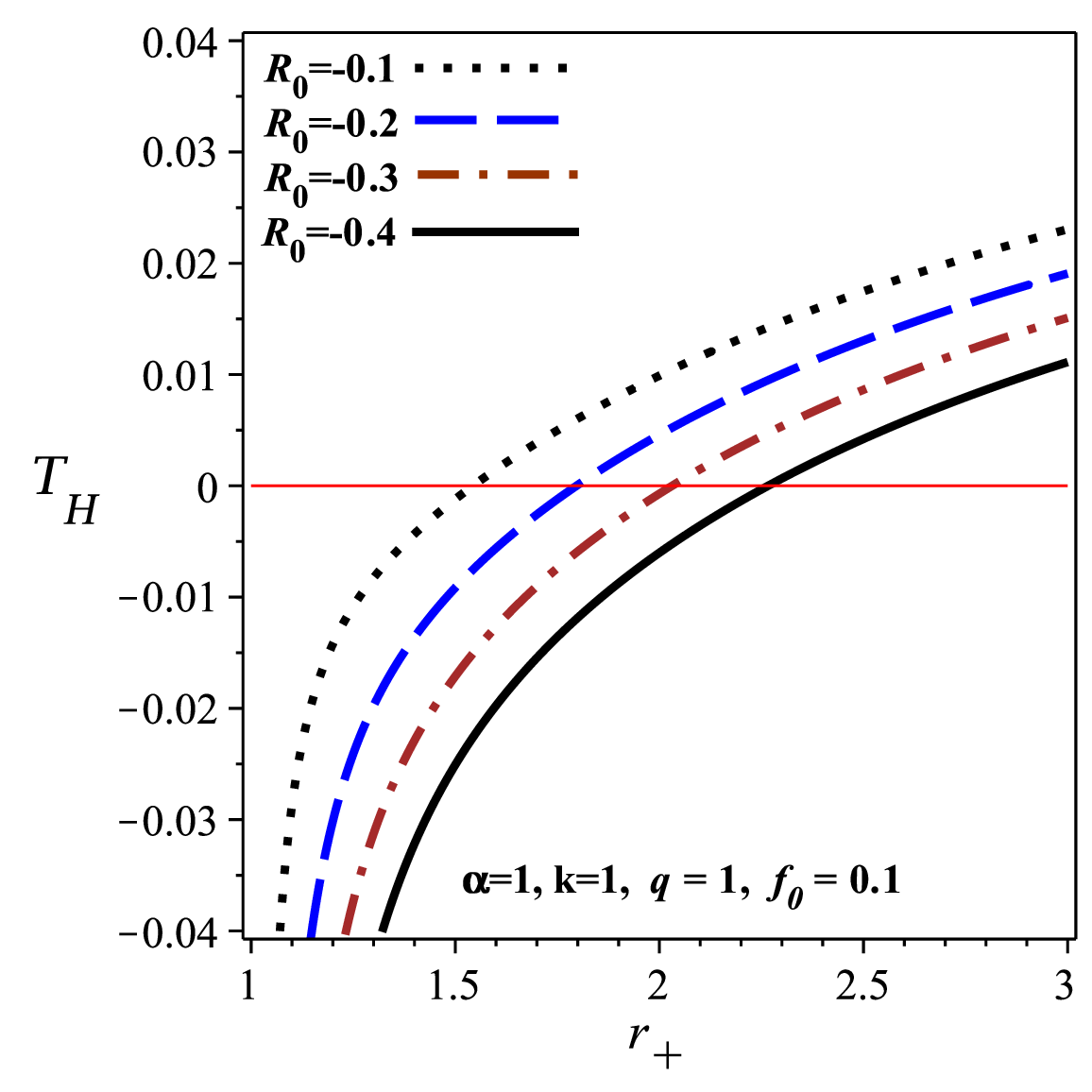} \includegraphics[width=0.32\linewidth]{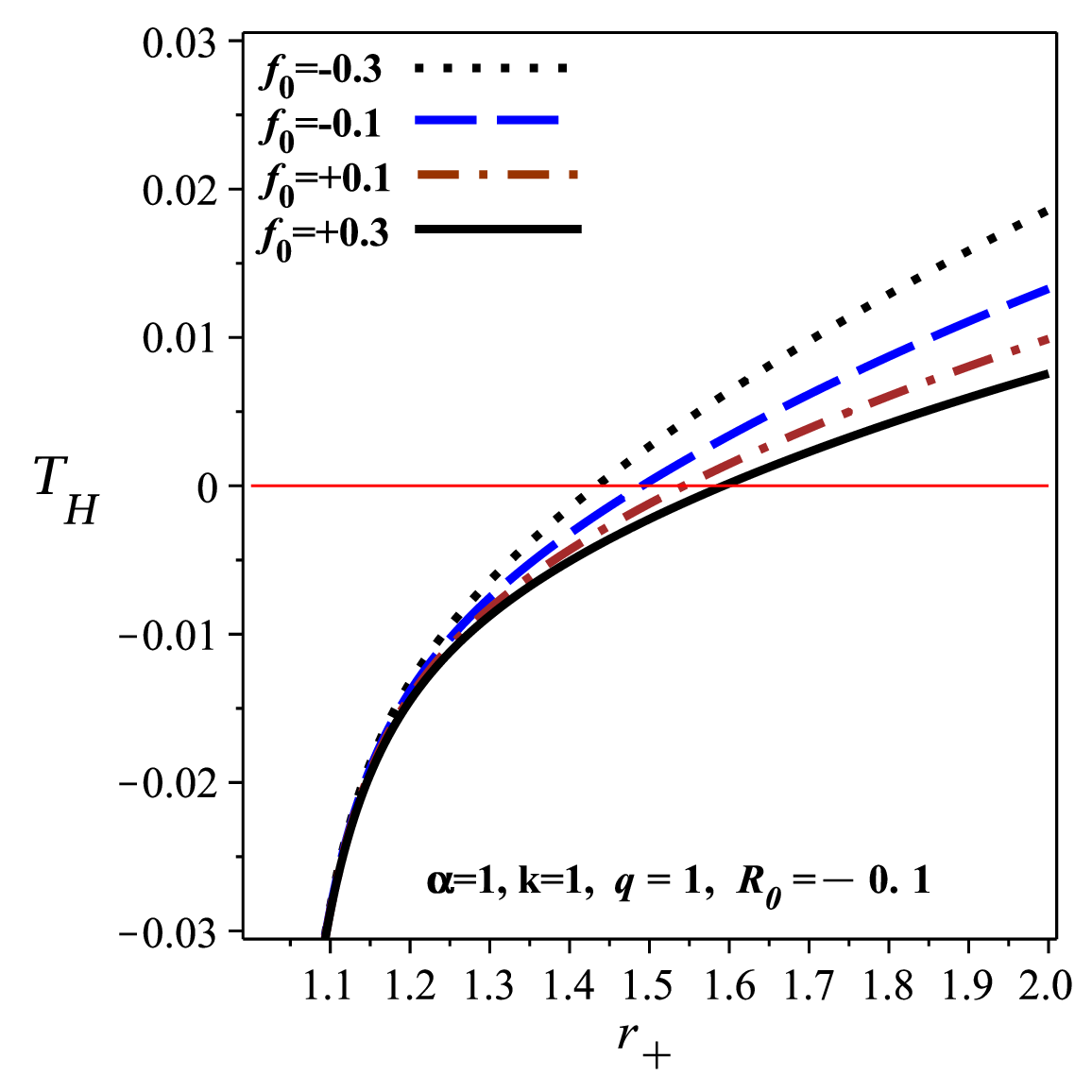} \newline
\caption{$T_{H}$ versus $r_{+}$ for different values of the parameters.}
\label{Fig2}
\end{figure}

To see the effects of different parameters such as $q$, $R_{0}$ and $%
f_{R_{0}}$ on the temperature (\ref{TbtzFR}), we plot it in Fig. \ref{Fig2}.
As one can see, there is a root in which the temperature is negative
(positive) before (after) this point. This root changes by varying the
parameters of the black hole. The effect of electrical charge shows that the
higher-charged black holes have a large area of positive temperature (see
the left panel in Fig. \ref{Fig2}). The effect of scalar curvature reveals
that the negative temperature area of a black hole increases by increasing
the value of $\left\vert R_{0}\right\vert $\ (see the middle panel in Fig. %
\ref{Fig2}). The effect of $F(R)$\ gravity indicates that the positive
temperature area of the black hole decreases by increasing the value of $%
f_{0}$\ (see the right panel in Fig. \ref{Fig2}). However, there is the same
behavior for the temperature of these black holes. Indeed, the large black
holes have a positive temperature.

\section{Conclusions}

The exact (un)charged accelerating black hole solutions in three-dimensional
spacetime were extracted by considering $F(R)$ gravity in the (absence)
presence of nonlinear electrodynamics. Our results showed that
higher-charged accelerating black holes with massive mass have large roots,
provided the maximum mass of these black holes could not be more than a
critical value (i.e., $m<m_{critical}$). For $m>m_{critical}$, the black
holes had two real roots, which might be related to the event and the
cosmological horizons. Moreover, there was the same behavior for parameters
of $\theta $ and $\alpha $. In other words, there were two roots for the
obtained solutions when $\theta >\theta _{critical}$, and $\alpha
>\alpha_{critical}$. The real root of the charged accelerating black holes
decreased by increasing the parameter of $F(R)$ gravity ($f_{0}$) and the
absolute value of the Ricci scalar $\left\vert R_{0}\right\vert $.

Another result was related to the obtained asymptotical behavior of these
black holes. It was not precisely asymptotically AdS and depended on
different parameters of gravity.

We obtained the temperature of three-dimensional (un)charged accelerating
AdS black holes in $F(R)$ gravity. For uncharged cases, the temperature was
always positive (negative) when $r_{+}<\frac{1}{\alpha }$ ($r_{+}>\frac{1}{%
\alpha }$). For charged cases, there was one real root for the temperature,
in which the temperature was negative (positive) before (after) this root.
Indeed, the small black holes had a negative temperature, but the large
black holes were physical objects. We evaluated the effects of various
parameters on the temperature of these black holes that were briefly; i) the
higher-charged accelerating BTZ black holes had a large area of the positive
temperature. ii) the effect of scalar curvature indicated that the positive
temperature area of the black hole increases by decreasing the value of $%
\left\vert R_{0}\right\vert $. iii) the effect of $F(R)$ gravity showed that
the negative temperature area of the black hole increases by increasing the
value of $f_{0}$.

\begin{acknowledgements}
	
We would like to thank the referee for giving valuable comments to improve this manuscript. B. Eslam Panah thanks University of Mazandaran.
		
\end{acknowledgements}

\section*{Data Availability Statement}

No data associated in the manuscript

\end{document}